\documentclass[preprint,showpacs,preprintnumbers,amssymb,aps,nofootinbib]{revtex4}
\usepackage{amsfonts,color,graphicx,epsfig,graphicx,amsmath,multirow,slashed}

\begin{document}

\title{$\Upsilon$ electroproduction at HERA, EIC, and LHeC within the nonrelativistic QCD framework}

\author{Zhan Sun}
\email{zhansun@cqu.edu.cn}

\affiliation{
\footnotesize
Department of Physics, Guizhou Minzu University, Guiyang 550025, People's Republic of China. \\
}

\date{\today}

\begin{abstract}

Based on the nonrelativistic QCD framework, we study the $\Upsilon$ production in semi-inclusive deep inelastic electron-proton scattering (SIDIS) at HERA, EIC, and LHeC, with the main aim of assessing the viability of observing $\Upsilon$ electroproduction at the three colliders. The color-octet (CO) contributions are found to have a crucial effect on both the integrated and differential cross sections, serving to further establish the significance of the CO mechanism. By setting the kinematic cuts to $p_{t,\Upsilon}^{\star 2}>1~\textrm{GeV}^{2}$, $W>50$ GeV, $2<Q^2<100~\textrm{GeV}^2$, and $0.3<z<0.9$, only a few electroproduced $\Upsilon$ events can be generated at HERA, partially accounting for its lack of measured $\Upsilon$ electroproduction. However, under the same cut conditions, the EIC and LHeC can accumulate about $8.7\times10^{2}$ and $3.7\times10^{4}$ reconstructed $\Upsilon$ events in one operation year, respectively, which manifestly indicates the prospect of detecting the $\Upsilon$-related SIDIS processes at the two forthcoming $ep(eA)$ colliders.


\end{abstract}

\maketitle

\section{Introduction}{\label{intro}}

The heavy quarkonium related production processes, which involve both perturbative (production of a heavy-quark pair) and nonperturbative (evolution of the heavy-quark pair into the physical quarkonium) aspects of QCD, have always been an area of great interest in high-energy physics. In addition to inclusive hadro- and photoproduction where the validity of the nonrelativistic QCD (NRQCD) factorization \cite{Bodwin:1994jh} has been proved by the success of the results it predicts, electroproduction in deeply inelastic lepton-hadron scattering (DIS), commonly viewed as the cleanest probe of partonic behavior in protons and nuclei, is another process of particular importance for studying the properties of heavy quarkonium. In contrast to inclusive photoproduction where the virtuality ($Q^2$) of the photon emitted from the initial electron is constrained to a very small value, e.g., $Q^2<1~\textrm{GeV}^2$, the DIS process allows for a much larger $Q^2$, which may make the perturbative calculations work better. An evident example is that the next-to-leading-order QCD correction to $J/\psi$ production in semi-inclusive DIS (SIDIS) exhibits a decent convergence \cite{Sun:2017wxk}, significantly better than the cases of hadro- and photoproduction \cite{Artoisenet:2008fc,Gong:2008sn,Lansberg:2008gk,Kramer:1995nb,Artoisenet:2009xh,Chang:2009uj}. Meanwhile, the large $Q^2$ will make the resolved photon effects less important than in the photoproduction case \cite{Adloff:2002ey} due to the rapid decrease in the probability of a photon to be resolved towards higher $Q^{2}$. Furthermore, the background from diffractive production is also expected to decrease faster with $Q^2$ than in the case of photoproduction \cite{Adloff:2002ey}. On the experimental side, the data given by Tevatron, LHC, etc., are mainly correlated to the quarkonium transverse momentum ($p_t$); however, as for the SIDIS process, many more varieties of physical observables can be measured, e.g., $p_t$, $p_t^{\star}$, $Y(\textrm{rapidity})$, $Y^{\star}$, $W$, $z$, and $Q^{2}$, which can aid in understanding the heavy-quarkonium production mechanism. Throughout the paper we employ the superscript $``\star"$ to denote the measured quantities in the center-of-mass (CM) frame of $\gamma^{*}p$. Moreover, the distinct signature of the scattered electron in the final state makes the progress easier to detect.

The electron-proton ($ep$) collider HERA released a large amount of data related to the $J/\psi$ production in SIDIS \cite{Adloff:2002ey,Adloff:1999zs,Chekanov:2005cf,Aaron:2010gz}, which apparently exceed the predictions given by the color-singlet mechanism (CSM) but coincide with the NRQCD ones \cite{Sun:2017wxk,Baier:1981zz,Korner:1982fm,Guillet:1987xr,Merabet:1994sm,Fleming:1997fq,Yuan:2000cn,Kniehl:2001tk,Sun:2017nly,Zhang:2017dia,Brambilla:2004wf,Brambilla:2010cs,Lansberg:2019adr,Zhang:2019ecf,Boer:2020bbd}. Despite the fruitful $J/\psi$ measurements, as a result of the low luminosity ($\sim 10^{31}~\textrm{cm}^{-2}~\textrm{s}^{-1} $) HERA has so far not observed a significant signal of $\Upsilon$ production in SIDIS; theoretically, far too little attention has been paid to this topic. In comparison with $J/\psi$, $\Upsilon$ has its own conspicuous advantages. First, the large mass of the $b$ quark makes both the typical coupling constant ($\alpha_s$) and relative velocity ($v$) smaller than in the $J/\psi$ case, in general resulting in a better convergent perturbative series over the expansion of $\alpha_s$ and $v^2$. Second, $\Upsilon$ can also be straightforwardly tagged by searching for its decays into a lepton pair; moreover, the large $b$-quark mass will further make these decay products more energetic and thereby more easily detectable. Additionally, there is no $b$-hadron feed-down contributing to $\Upsilon$ production. Thus, as the most studied $b\bar{b}$ meson in the bottomonium family, $\Upsilon$ may provide an even better opportunity to apply the NRQCD framework than the $J/\psi$ case. In combination with the aforementioned DIS-process benefits, $\Upsilon$ electroproduction is expected to provide an ideal laboratory for the study of heavy quarkonium, meriting a separate investigation.

As previously stated, the low luminosity of HERA highly suppressed its ability to search for $\Upsilon$ production in SIDIS, subsequently creating obstacles for the corresponding theoretical studies. Fortunately, the forthcoming Large Hadron-Electron Collider (LHeC) \cite{AbelleiraFernandez:2012cc,AbelleiraFernandez:2012ni} and the Electron-Ion Collider (EIC) \cite{Accardi:2012qut}, which exceed HERA's luminosity by 2-3 orders of magnitude, bring great opportunities to fulfill the observation of $\Upsilon$ electroproduction. The LHeC, designed as a second (first) generation DIS $ep~(eA)$ collider by CERN, takes advantage of a newly built electron beam of 60 GeV (and up to possibly 140 GeV) which will collide with the intense, high-energy hadron beams of the LHC. It has a unique DIS physics programme which can be pursued with unprecedented precision over a widely extended kinematic range. The EIC proposed by the Brookhaven National Laboratory is designed to utilize a new electron beam facility based on an energy recovery linac to be built inside the RHIC tunnel to collide with RHIC's existing high-energy proton and nuclear beams, with the CM energy varying from $20-140$ GeV. These high luminosities provide LHeC and EIC\footnote{ The beneficial features of EIC for studying the heavy-quarkonium production have been exemplified by the inclusive $J/\psi$ photoproductions \cite{Flore:2020jau,Qiu:2020xum}.} with an excellent ability to perform a multitude of crucial DIS measurements. In light of these outstanding merits, in this paper wel investigate and compare $\Upsilon$ production in $ep$ SIDIS at LHeC, EIC, and HERA, in order to pave the way for future comparisons between measurements and corresponding theoretical predictions.

The rest of the paper is organized as follows. In Sec. II we talk about the choice of the parameters used in our calculations and present the phenomenological results. In Section III we provide a summary.
%----------------------------------------------------------------------
\section{Framework and Phenomenological results}
\subsection{Framework}
\begin{figure}
\includegraphics[width=0.5\textwidth]{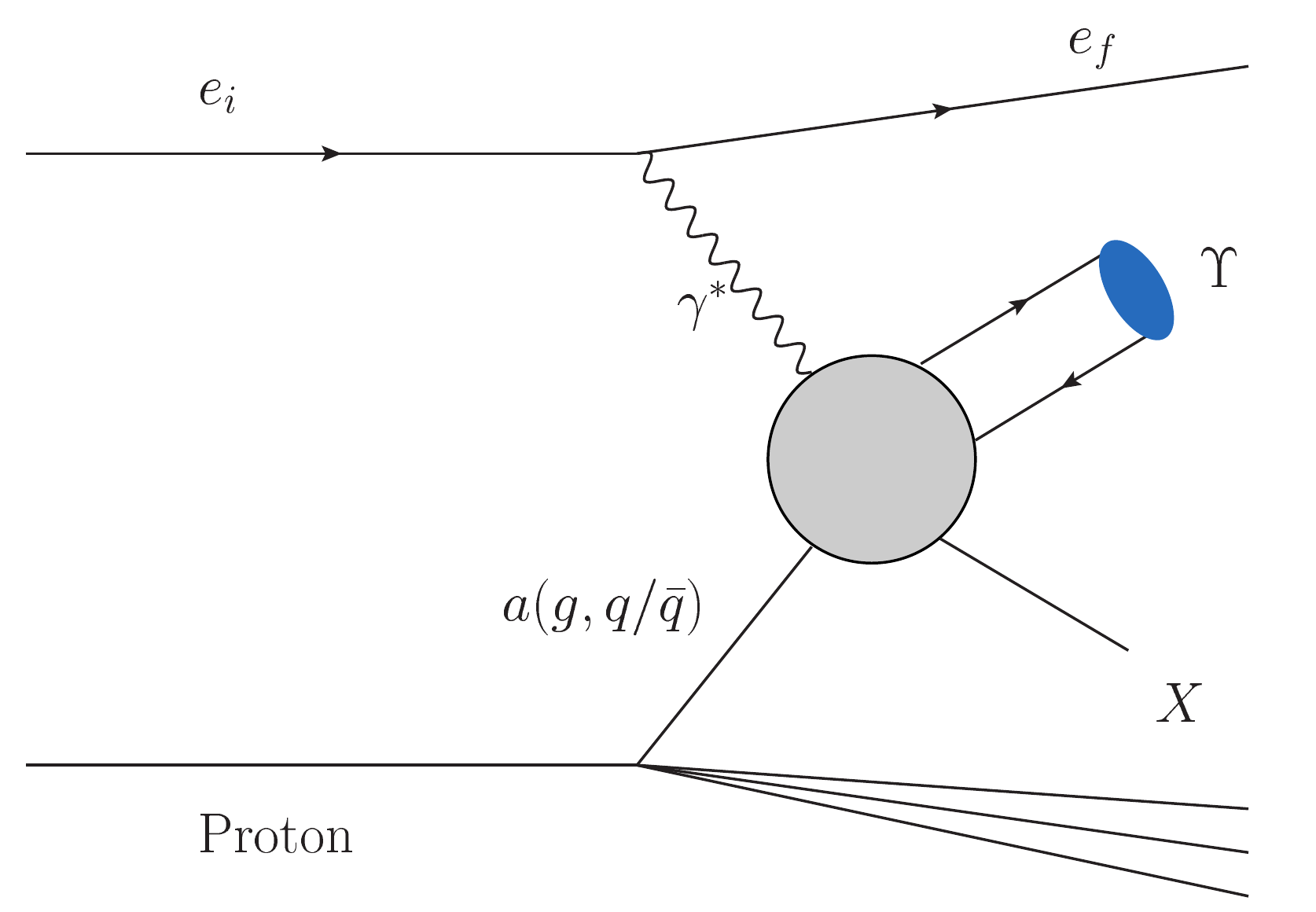}
\caption{\label{fig:Feyn}
The illustrative diagram for the $\Upsilon$ production via $e+p \to e+\Upsilon+X$. The subscript $``q"$ represents the light quarks $(u,d,s)$.}
\end{figure}

The procedures for dealing with the $J/\psi$ (or $\Xi_{cc}$) related SIDIS processes have been described in detail in our previous papers \cite{Sun:2017wxk,Sun:2017nly,Zhang:2017dia,Sun:2020mvl}, and thus we will directly adopt the same calculation formalism to approach $\Upsilon$ production via SIDIS, which is schematically depicted in Fig. \ref{fig:Feyn}. 

The parameter choices used in the calculations are listed below:
\begin{itemize}
\item[1)]
The respective collision energy of HERA, EIC, and LHeC are summarized in Table \ref{energy}, where $E_e$ and $E_p$ refer to the electron and proton beam energies, respectively, and $\sqrt{S}$ is the CM energy.
\begin{table*}[htb]
\caption{The beam energies and CM energies of HERA, EIC, and LHeC in units of GeV.}
\label{energy}
\begin{tabular}{lccccccc}
\hline\hline
$~~~$ & $~~~E_e~~~$ & $~~~E_p~~~$ & $~~~\sqrt{S}~~~$\\ \hline
HERA & $27.5$ & $920$ & $318$\\
EIC & $21$ & $100$ & $91.65$\\
LHeC & $60$ & $7000$ & $1296$\\ \hline\hline
\end{tabular}
\end{table*}
\item[2)]
Seeing that no evident signal of $\Upsilon$ production via SIDIS has been obtained so far, we will take $J/\psi$ electroproduction at HERA for reference to assume the kinematic cuts for the $\Upsilon$ case. From \cite{Adloff:2002ey,Adloff:1999zs,Chekanov:2005cf,Aaron:2010gz} we know that the $J/\psi$ related SIDIS processes at HERA mainly cover the ranges $p_{t,J/\psi}^{\star2}>1~\textrm{GeV}^2$, $0.3<z<0.9$, $2<Q^2<100~\textrm{GeV}^2$, and $W>50$ GeV. The cuts $p_{t,J/\psi}^{\star2}>1~\textrm{GeV}^2$ together with $z<0.9$ are applied to suppress the diffractive effects; $z>0.3$ is taken to exclude the contributions of the $b$ hadron decay and the resolved photon. With respect to $\Upsilon$ electroproduction, the situation is analogous to the $J/\psi$ case, and hence in our calculation we take the following cut conditions:
\begin{eqnarray}
p_{t,\Upsilon}^{\star2}&>&1~\textrm{GeV}^2,~W>50~\textrm{GeV}, \nonumber \\
2&<&Q^2<100~\textrm{GeV}^2, \nonumber \\
0.3&<&z<0.9.
\label{cuts}
\end{eqnarray}
\item[3)]
The $b$-quark mass is taken as $m_b=M_{\Upsilon}/2=4.75$ GeV \cite{Han:2014kxa} and the fine- structure constant is $\alpha=1/137.$
\item[4)]
The factorization and renormalization scales are chosen to be $\mu_f=\mu_r=\xi\sqrt{Q^2+M^2_{\Upsilon}}$ with $\xi$ varying  between $1/2$ and $2$ around the default value 1.
\item[5)]
To determine the NRQCD LDMEs, we take advantage of the two linear combinations given by Ref. \cite{Han:2014kxa}, \footnote{In our calculation, we concentrate on direct $\Upsilon(1S)$ production, completely excluding the feed-down contributions via the higher excited states.}
\begin{eqnarray}
M_{0,r_0}^{\Upsilon}&=&\langle\mathcal{O} ^{\Upsilon}(^1S_0^{[8]})\rangle+\frac{r_0}{m_b^2}\langle\mathcal{O} ^{\Upsilon}(^3P_0^{[8]})\rangle, \nonumber \\
M_{1,r_1}^{\Upsilon}&=&\langle\mathcal{O} ^{\Upsilon}(^3S_1^{[8]})\rangle+\frac{r_1}{m_b^2}\langle\mathcal{O} ^{\Upsilon}(^3P_0^{[8]})\rangle,
\end{eqnarray}
where $r_0=3.8$, $r_1=-0.52$, $M_{0,r_0}^{\Upsilon}=13.7 \times 10^{-2}~\textrm{GeV}^{3}$, and $M_{1,r_1}^{\Upsilon}=1.17 \times 10^{-2}~\textrm{GeV}^{3}$. We set $\langle\mathcal{O} ^{\Upsilon}(^1S_0^{[8]})\rangle$ to $\zeta M_{0,r_0}^{\Upsilon}$ and correspondingly $\langle\mathcal{O} ^{\Upsilon}(^3P_0^{[8]})\rangle=(1-\zeta)(m_b^{2}/r_0)M_{0,r_0}^{\Upsilon}$, and vary $\zeta$ between 0 and 1 around the default value $1/2$. The CS LDME $\langle\mathcal{O} ^{\Upsilon}(^3S_1^{[1]})\rangle$ is related to the $S$-wave function at the origin by the following formula:
\begin{eqnarray}
\frac{\langle\mathcal{O} ^{\Upsilon}(^3S_1^{[1]})\rangle}{6N_c}=\frac{1}{4\pi}|R_{\Upsilon}(0)^2|,
\end{eqnarray}
with $|R_{\Upsilon}(0)^2|=6.477~\textrm{GeV}^3$ \cite{Eichten:1995ch}.
\end{itemize}
%----------------------------------------------------------------------
\subsection{Phenomenological results}

\begin{table*}[htb]
\caption{The integrated cross sections (in units of pb) of $\Upsilon$ electroproducion in correspondence to $^1S_0^{[8]}$,  $^3S_1^{[8]}$, $^3P_J^{[8]}$, and $^3S_1^{[1]}$, respectively. $p_{t,\Upsilon}^{\star 2}>1~\textrm{GeV}^{2}$, $W>50$ GeV, $2<Q^2<100~\textrm{GeV}^2$, and $0.3<z<0.9$.}
\label{integrated cross section}
\begin{tabular}{lcccccccc}
\hline\hline
$\sigma[n]$ & $~n=^1S_0^{[8]}~$ & $~n=^3S_1^{[8]}~$ & $~n=^3P_J^{[8]}~$ & $~n=^3S_1^{[1]}~$\\ \hline
HERA & $0.342$ & $0.009$ & $0.448$ & $1.065$\\
EIC & $0.033$ & $0.001$ & $0.043$ & $0.097$\\
LHeC & $1.691$ & $0.035$ & $2.200$ & $5.362$\\ \hline\hline
\end{tabular}
\end{table*}

The integrated cross sections of $\Upsilon$ production in SIDIS corresponding to different Fock states are listed in Table \ref{integrated cross section}. As can be seen from this table, including the CO ($^1S_0^{[8]}$, $^3S_1^{[8]}$, and $^3P_J^{[8]}$) contributions can to a large extent enhance the CS ($^3S_1^{[1]}$) predictions, by about $75-80\%$; among the CO channels, the contributions of $^1S_0^{[8]}$ and $^3P_J^{[8]}$ play the leading role, absolutely dominating over that of the $^3S_1^{[8]}$ configuration. 

By summing up the CS and CO contributions, we finally get for the NRQCD predictions:
\begin{eqnarray}
\sigma_{\textrm{HERA}}&=&1.864^{+0.695+0.109}_{-0.489-0.109}~\textrm{pb}, \nonumber \\
\sigma_{\textrm{EIC}}&=&0.174^{+0.099+0.010}_{-0.058-0.010}~\textrm{pb}, \nonumber \\
\sigma_{\textrm{LHeC}}&=&9.288^{+1.743+0.525}_{-1.620-0.525}~\textrm{pb},\label{ICS}
\end{eqnarray}
where the two columns of uncertainties are caused by varying $\xi$ from 1/2 to 2 around 1 and varying $\zeta$ from 0 to 1 around 1/2, respectively. For HERA and EIC, which are equipped with low CM energies, the ambiguity of the choice of the renormalization and factorization scales serves as the most important source of the theoretical uncertainties. For example, halving (doubling) the default values of $\mu_r$ and $\mu_f$  simultaneously will enlarge (diminish) the cross sections by about $57\% (34\%)$ for EIC, and $37\% (26\%)$ for HERA. Regarding LHeC, which runs at a much higher collision energy, varying $\xi$ from 1 to 1/2(2) brings about a $19\% (17\%)$ fluctuation of the integrated cross section. Varying $\zeta$ from 0 to 1 around 1/2 only changes the results by about $\pm 5\%$.

As pointed out in Sec. \ref{intro}, HERA has not yet obtained any measurement on $\Upsilon$ production in SIDIS. To clarify this issue quantitatively, we use the numerical result in Eq. (\ref{ICS}) to estimate the possible number of $\Upsilon$ electroproduction events at HERA. According to the integrated luminosity ($63~\textrm{pb}^{-1}$) corresponding to $E_{p}=920$ GeV during 1997$-$2000 \cite{Adloff:2002ey}, only about five electroproduced $\Upsilon$ events\footnote{By taking into account the detection efficiency, the event number should be further reduced.} [which are established by searching for $\Upsilon \to l^{+}l^{-}$ ($\simeq 5\%$)] can be accumulated, thereby making the detection extremely difficult. 

By assuming $\mathcal{L}_{\textrm{EIC}}=10^{34}~\textrm{cm}^{-2}~\textrm{s}^{-1}$ \cite{Accardi:2012qut} and $\mathcal{L}_{\textrm{LHeC}}=0.8 \times 10^{34}~\textrm{cm}^{-2}~\textrm{s}^{-1}$ \cite{AbelleiraFernandez:2012cc,AbelleiraFernandez:2012ni,Tanabashi:2018oca}\footnote{As reported in Ref. \cite{Tanabashi:2018oca}, by the upgraded proton beam parameters of the HL-LHC, the up-to-date luminosity of LHeC could be improved to be $\mathcal{L}_{\textrm{LHeC}}=0.8 \times 10^{34}\textrm{cm}^{-2}\textrm{s}^{-1}$.}, and a detection efficiency of $100\%$, about $3.7 \times 10^{4}$ and $8.7 \times 10^{2}$ reconstructed $\Upsilon$ events are estimated to be generated in one operation year ($10^{7}$ seconds running time\footnote{Approximately, 1 year $\approx \pi \times 10^{7}$ seconds, but it is common that a collider only operates about $1/\pi$ year, i.e., $10^{34}~\textrm{cm}^{-2}~\textrm{s}^{-1} \approx 10^{5}~\textrm{pb}^{-1}/\textrm{year}$.}) at the LHeC and EIC, respectively, which suggests that the two forthcoming $ep(eA)$ collider are capable of observing $\Upsilon$ production in SIDIS.

\begin{figure*}
\includegraphics[width=0.32\textwidth]{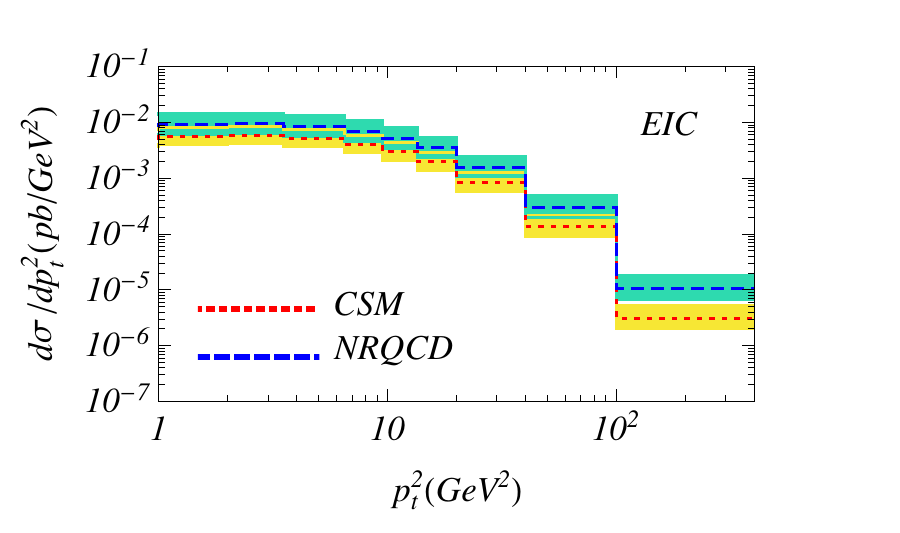}
\includegraphics[width=0.32\textwidth]{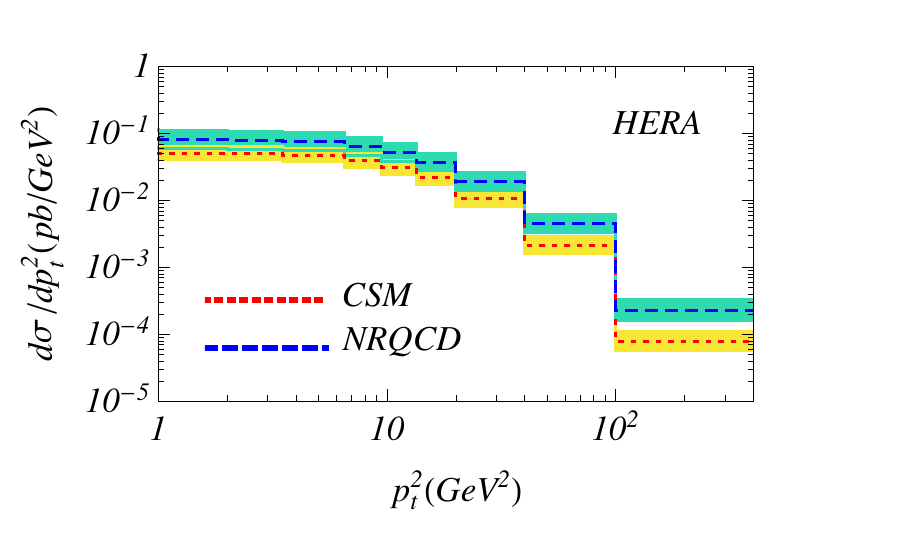}
\includegraphics[width=0.32\textwidth]{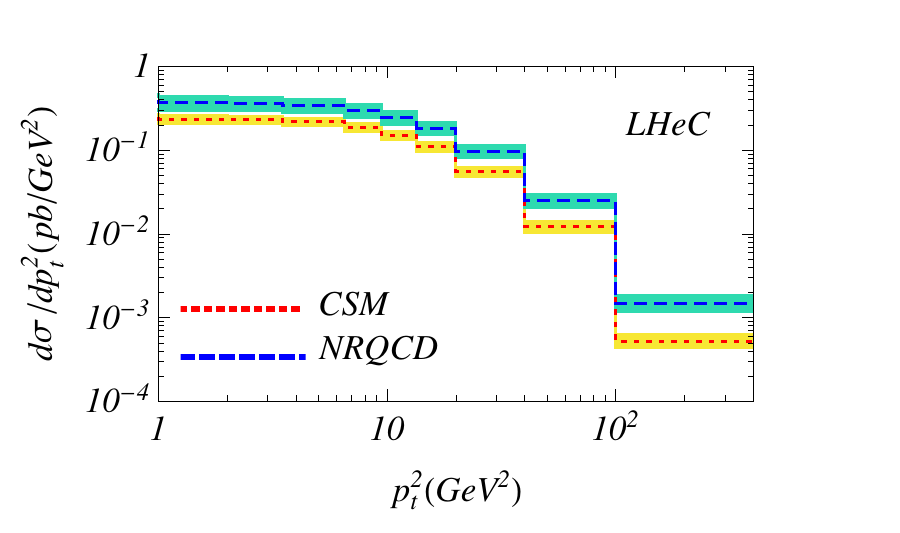}
\includegraphics[width=0.32\textwidth]{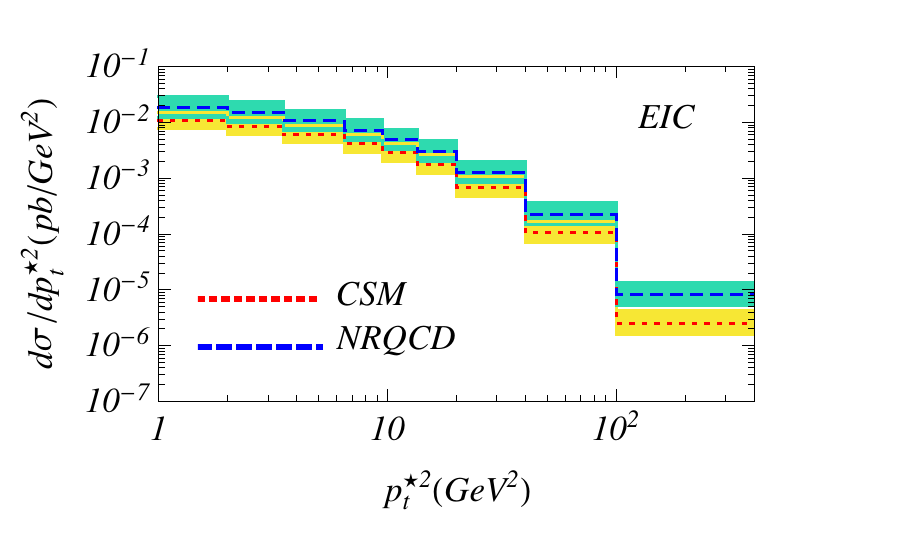}
\includegraphics[width=0.32\textwidth]{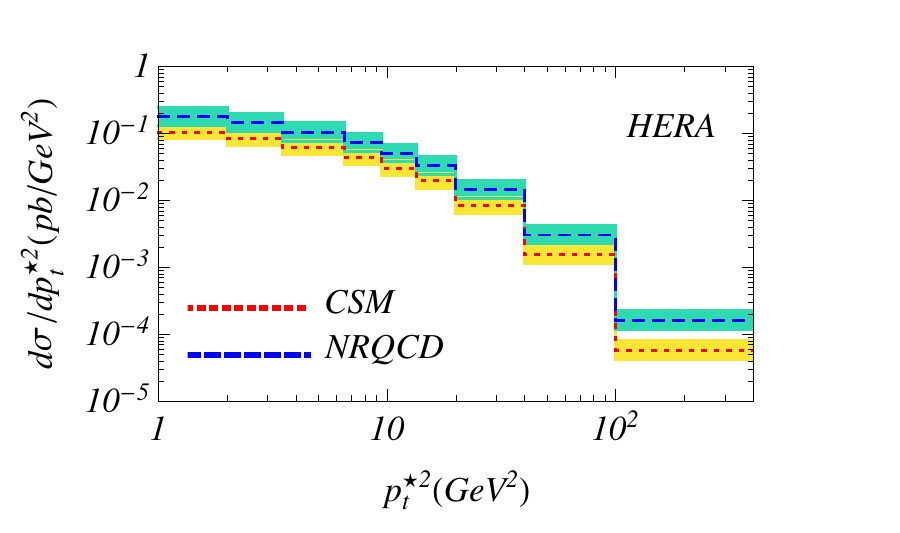}
\includegraphics[width=0.32\textwidth]{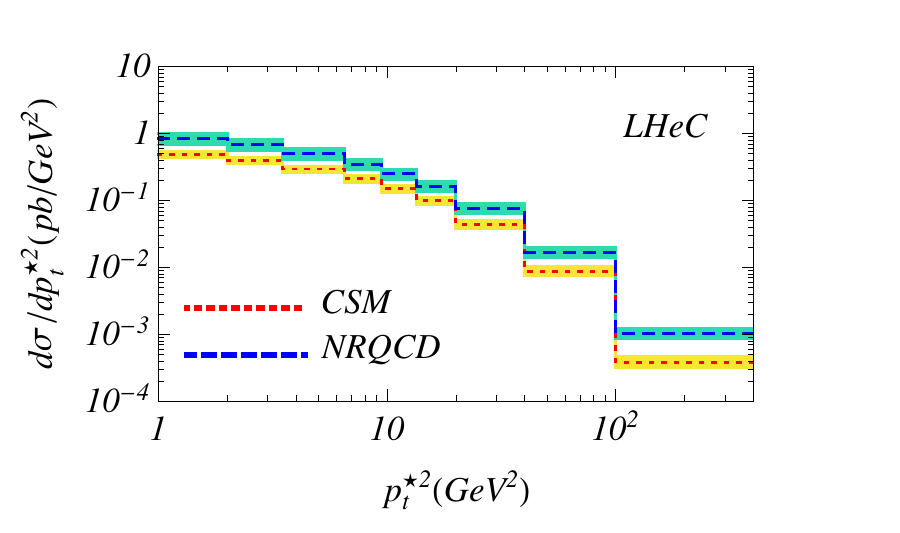}
\caption{\label{fig:pt}
The distributions of $p_t^2$ and $p_t^{\star2}$ under the kinematic cuts in Eq. (\ref{cuts}). The shaded bands are induced by the variations of $\xi$ and $\zeta$.}
\end{figure*}

\begin{figure*}
\includegraphics[width=0.32\textwidth]{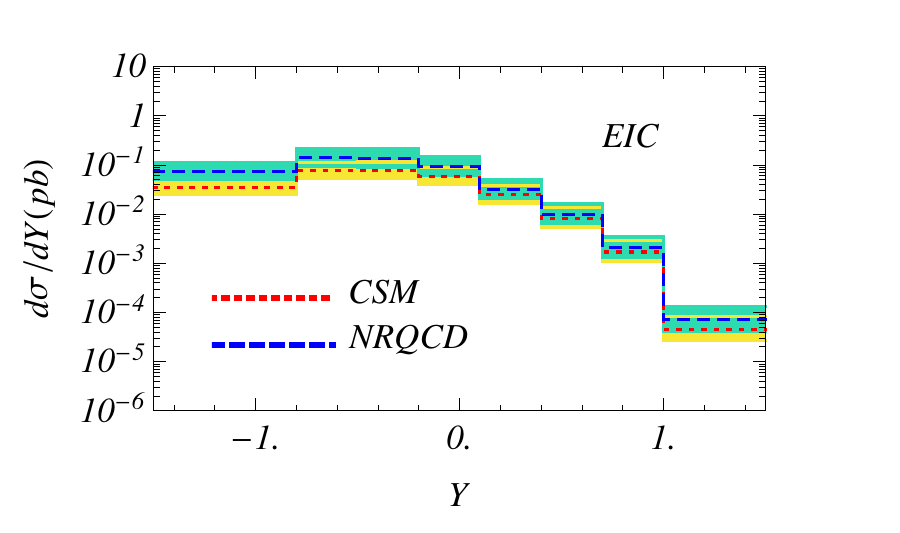}
\includegraphics[width=0.32\textwidth]{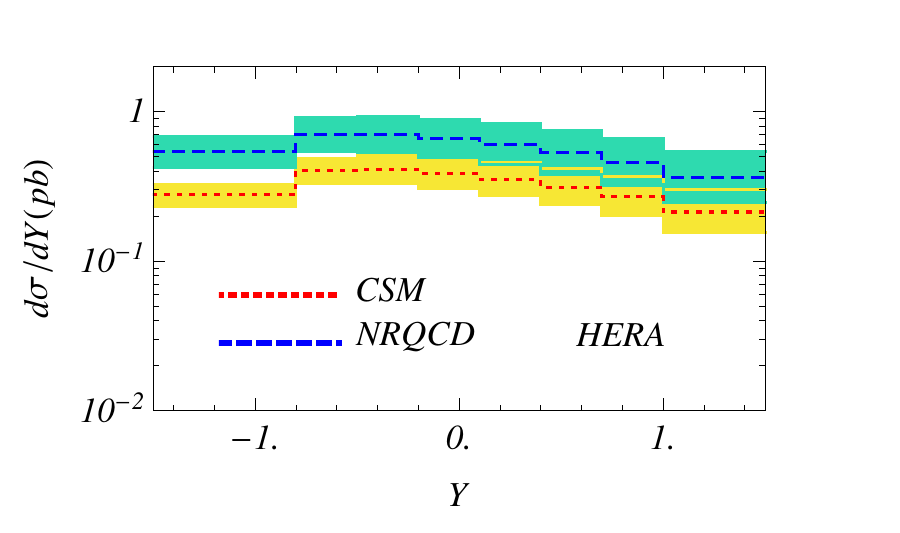}
\includegraphics[width=0.32\textwidth]{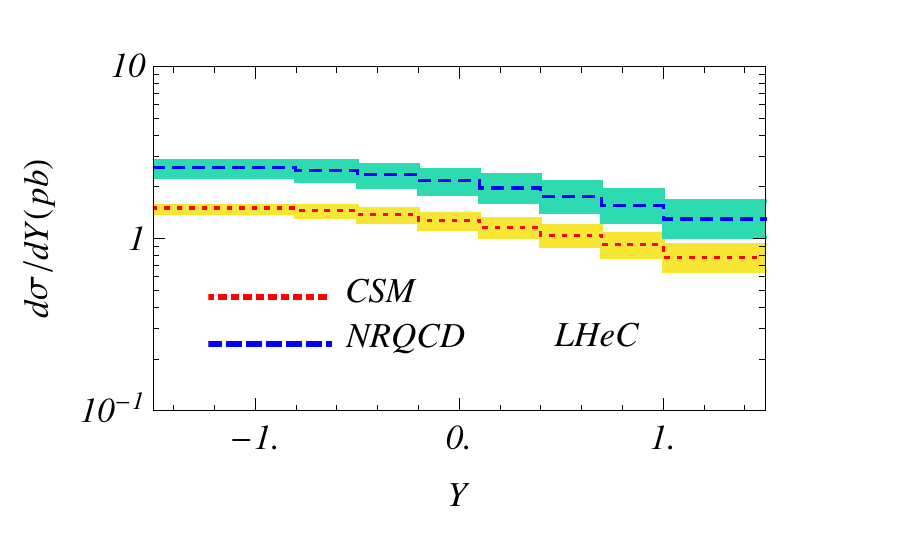}
\includegraphics[width=0.32\textwidth]{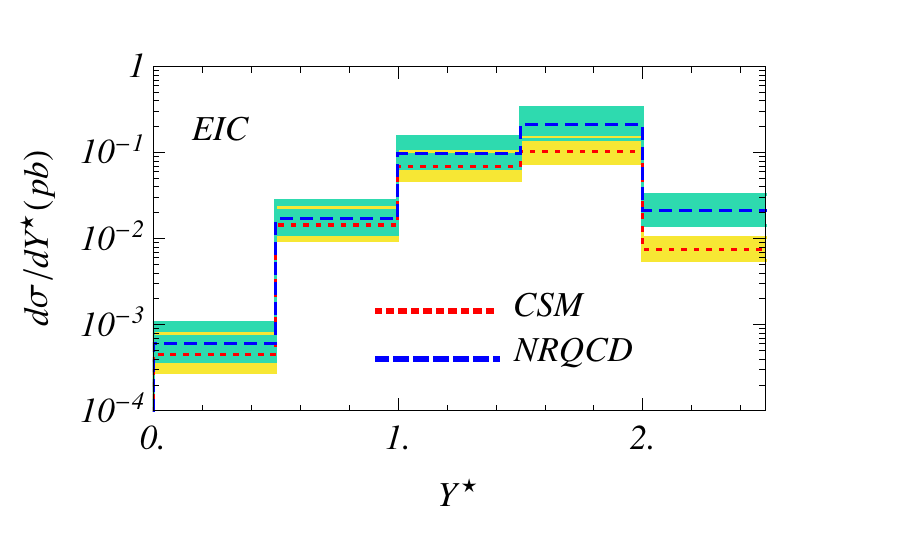}
\includegraphics[width=0.32\textwidth]{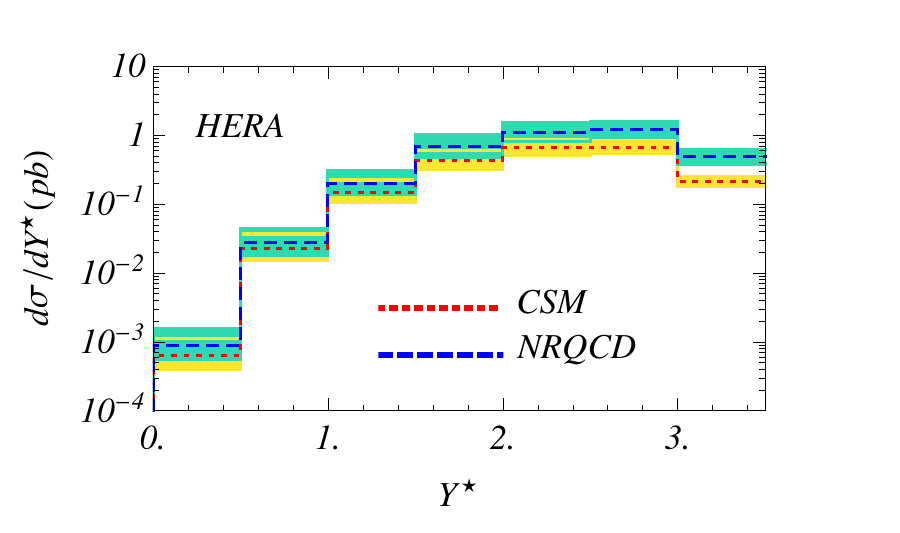}
\includegraphics[width=0.32\textwidth]{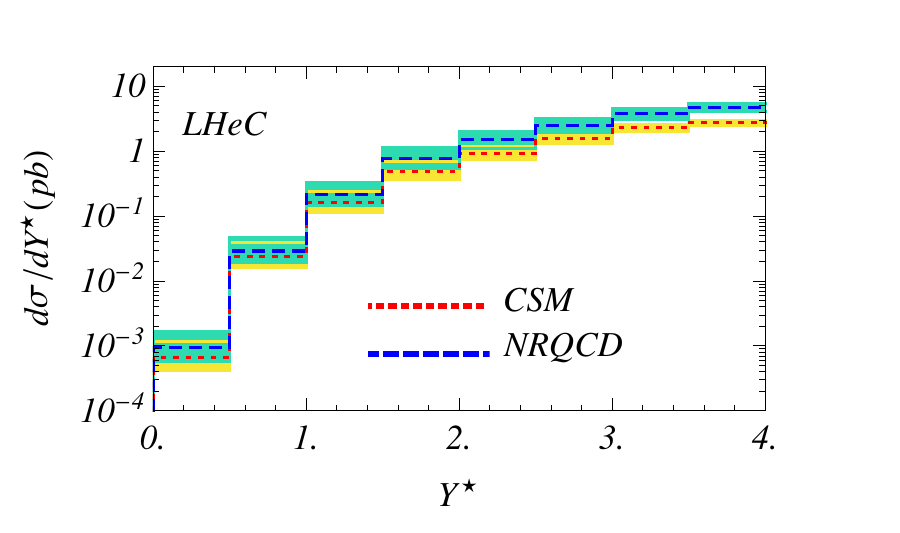}
\caption{\label{fig:y}
The distributions of $Y$ and $Y^{*}$ under the kinematic cuts in Eq. (\ref{cuts}). The shaded bands are induced by the variations of $\xi$ and $\zeta$.}
\end{figure*}

\begin{figure*}
\includegraphics[width=0.32\textwidth]{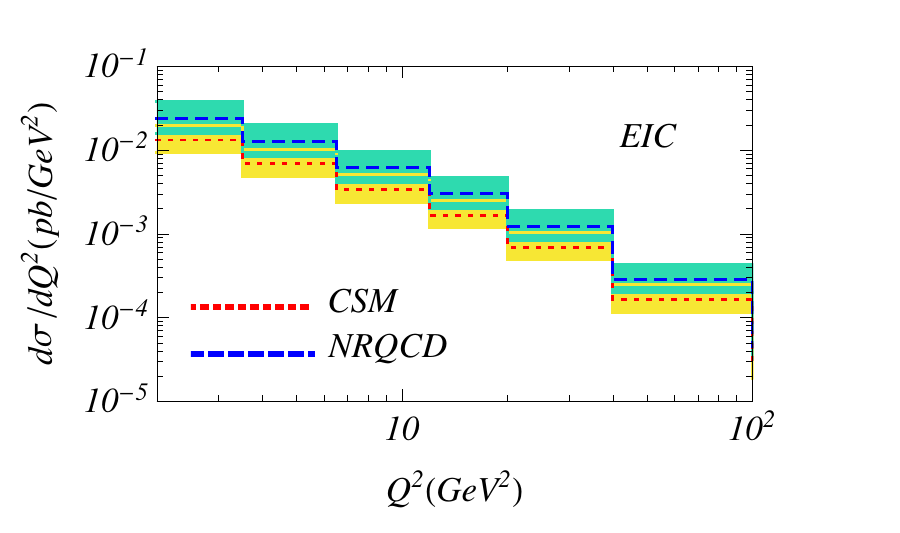}
\includegraphics[width=0.32\textwidth]{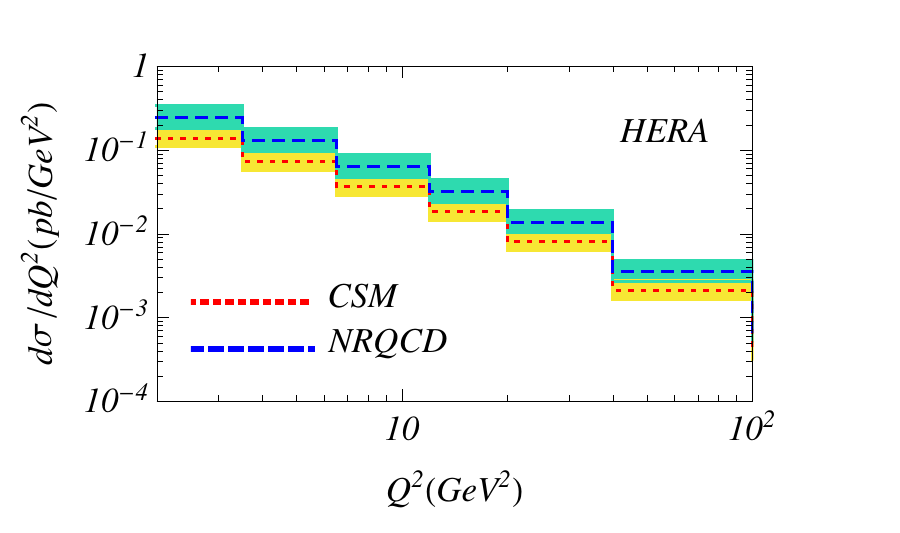}
\includegraphics[width=0.32\textwidth]{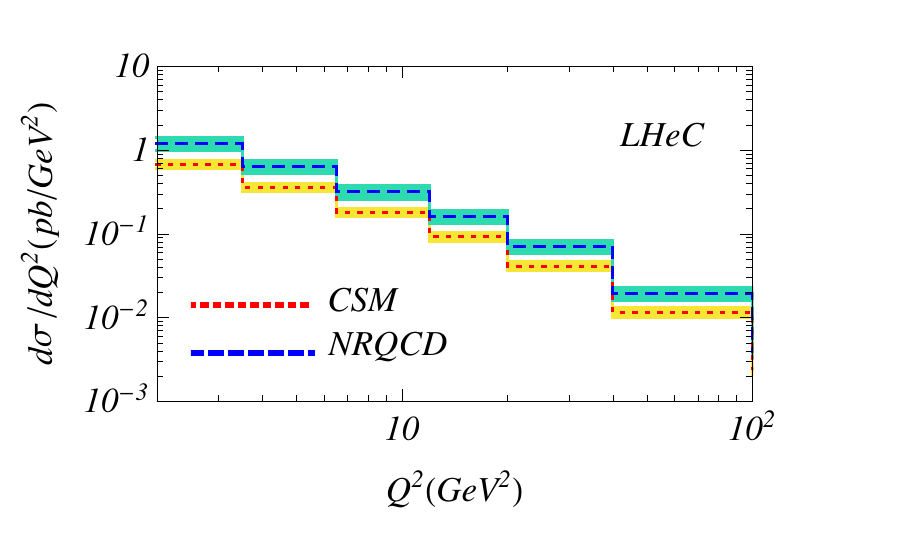}
\includegraphics[width=0.32\textwidth]{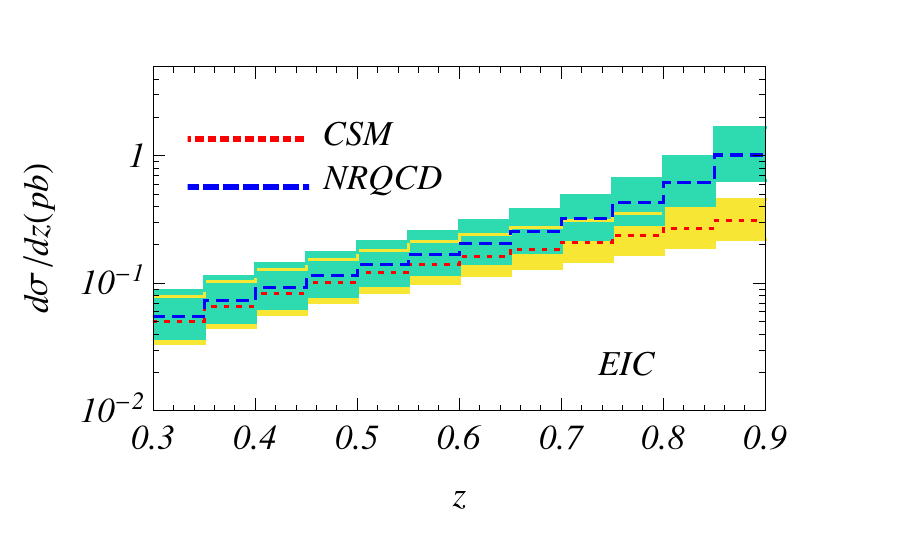}
\includegraphics[width=0.32\textwidth]{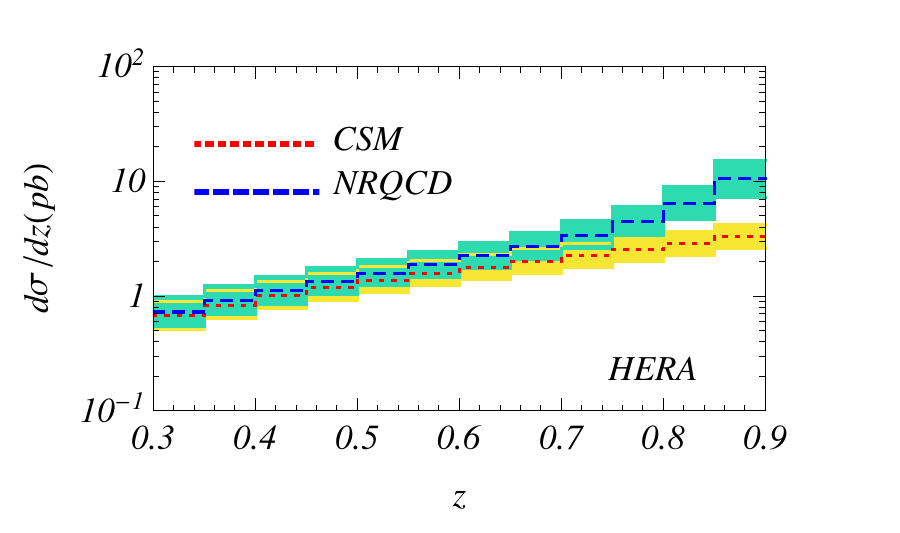}
\includegraphics[width=0.32\textwidth]{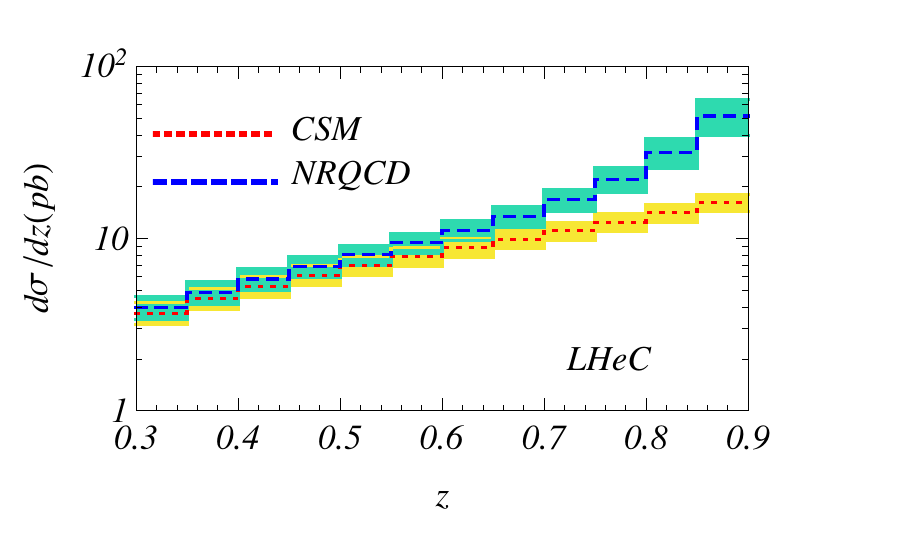}
\caption{\label{fig:q2z}
The distributions of $Q^2$ and $z$ under the kinematic cuts in Eq. (\ref{cuts}). The shaded bands are induced by the variations of $\xi$ and $\zeta$.}
\end{figure*}

The distributions of $p_{t}^{2}$, $p_{t}^{\star2}$, $Q^2$, $z$, $Y$, and $Y^{\star}$ are shown in Figs. \ref{fig:pt}, \ref{fig:y}, and \ref{fig:q2z}, where the NRQCD predictions are compared with the CS ones. Following the conventions of HERA, the forward direction of $Y^{\star}$ is defined as that of the incident virtual photon; $Y$ is taken to be positive in the direction of the incoming proton. In each diagram, the shaded bands denote the theoretical uncertainties stemming from the variations of $\xi$ and $\zeta$. From the figures we can see the following
\begin{itemize}
\item[1)]
As $p_{t}^{2}(p_{t}^{\star2})$ increases, the corresponding differential cross sections continuously decrease, with the CO contributions playing an increasingly crucial role. When $p_{t}^{2}(p_{t}^{\star2})<10~\textrm{GeV}^2$, the NRQCD predictions are less than 2 times larger than the CS ones; however, when $p_{t}^{2}(p_{t}^{\star2})$ is scattered around $100~\textrm{GeV}^2$, the former ones notably go beyond the latter ones.
\item[2)]
The differential cross sections as functions of $Y^{\star}$ exhibit serious asymmetry. Taking LHeC for example, the value of $d\sigma/dY^{\star}$ at $Y^{\star}=4$ is about 4 orders of magnitude in the excess of that at $Y^{*}=0$, which signifies that, in the $\gamma^{\star}p$ CM frame, $\Upsilon$ is much more likely to be produced in the direction of the virtual photon than in that of the incoming proton. From the three $Y^{\star}$-distribution figures in Fig. \ref{fig:y}, one can find that the available $Y^{\star}$ range will extend with the increment of the CM energy. $d\sigma/dY$ is also asymmetric, as is manifestly shown by the triple $Y$-distribution figures in Fig. \ref{fig:y}. This asymmetry tells us that, in the laboratory frame, more $\Upsilon$ events will be generated along the direction of the incident electron.
\item[3)]
With regard to the $Q^2$ distributions, the differential cross sections gradually fall off with increasing $Q^2$; the ratio of the NRQCD prediction to the CS one appears to be insensitive to $Q^{2}$. As for the $z$ distributions, when $z$ approaches 0.3 the CS contribution almost saturates the NRQCD prediction alone; however, towards higher $z$ values, the CO contribution rises sharply,\footnote{Near the endpoint region, $z \approx 1$, the perturbative and velocity expansions maybe break down, similar to the inclusive $J/\psi$ photo- and electroproductions. Resuming the series in $\alpha_s$ and $v$ perhaps to certain degree smear the CO predicted steep ascent of $d\sigma/dz$ at large $z$ values.} which can be primarily attributed to the $^1S_0^{[8]}$ and $^3P_J^{[8]}$ channels.
\end{itemize}

\begin{table*}[htb]
\caption{The integrated cross sections (in units of pb) of $\Upsilon$ electroproducion from the LDMEs of Refs. \cite{Feng:2015wka,Gong:2013qka,Sharma:2012dy}. $p_{t,\Upsilon}^{\star 2}>1~\textrm{GeV}^{2}$, $W>50$ GeV, $2<Q^2<100~\textrm{GeV}^2$, and $0.3<z<0.9$. }
\label{integrated cross section II}
\begin{tabular}{llccccccc}
\hline\hline
$~$ & $~$ & $~~~~~{\rm{Sharman}}$\cite{Sharma:2012dy} & $~~~~~{\rm{Gong}}$\cite{Gong:2013qka} & $~~~~~{\rm{Feng}}$\cite{Feng:2015wka}({\rm{Table~I}})\\ \hline
HERA & CSM & $1.263$ & $1.065$ & $1.065$\\
$~$ & NRQCD & $2.833$ & $1.454$ & $1.516$\\
EIC & CSM & $0.115$ & $0.097$ & $0.097$\\
$~$ & NRQCD & $0.266$ & $0.134$ & $0.140$\\
LHeC & CSM & $6.356$ & $5.362$ & $5.362$\\
$~$ & NRQCD & $14.06$ & $7.289$ & $7.594$\\ \hline\hline
\end{tabular}
\end{table*}

One of the crucial issues of NRQCD research is the determination of the LDMEs, of which there are several independent extractions, with different strategies having obtained different results. Therefore, to serve as a sound reference, we employ three other typical sets of these parameters \cite{Feng:2015wka,Gong:2013qka,Sharma:2012dy} to present our numerical results. From the data in Tale \ref{integrated cross section II}, one can find that the predicted integrated cross sections given by the LDMEs in Ref. \cite{Gong:2013qka} are approximately identical with those from the LDMEs of Ref. \cite{Feng:2015wka}, similar to the results in Eq. (\ref{ICS}). However, the large values of the CO LDMEs in Ref. \cite{Sharma:2012dy} significantly increase the CO contributions, subsequently making their NRQCD predictions about twice large than the ones from Refs. \cite{Han:2014kxa,Feng:2015wka,Gong:2013qka}.

To summarize, the CO configurations, especially $^1S_0^{[8]}$ and $^3P_J^{[8]}$, play a pivotal role in $\Upsilon$ production via SIDIS. The predicted remarkable discrepancies between the NRQCD predictions and the CS ones, which await future identification by LHeC and EIC can serve as useful evidence to favor or disfavor the CO processes.
%----------------------------------------------------------------------
\section{Summary}
In order to provide a deeper insight into heavy-quarkonium production, in this work we investigated the $\Upsilon$ production via the $ep$ SIDIS at HERA, EIC, and LHeC, within the NRQCD framework. We found that the CO states, especially $^1S_0^{[8]}$ and $^3P_J^{[8]}$, supply substantial contributions to $\Upsilon$ electroproduction, leading to distinct dissimilarities between the CS predictions and the NRQCD ones, which are beneficial to distinguish between the CS and CO mechanisms. Under the assumed constraints $p_{t,\Upsilon}^{\star 2}>1~\textrm{GeV}^{2}$, $W>50$ GeV, $2<Q^2<100~\textrm{GeV}^2$, and $0.3<z<0.9$, a rather small number of $\Upsilon$ electroproduction events can be gathered by HERA, which is responsible, in part, for the fact that it has not yet released any data of $\Upsilon$ production in SIDIS. However, as many as $3.7 \times 10^{4}$ and $8.7 \times 10^{2}$ electroproduced $\Upsilon$ events can be collected in one operation year at LHeC and EIC, respectively, clearly revealing the experimental prospect of observing $\Upsilon$ production via SIDIS at the two future $ep(eA)$ colliders.
%----------------------------------------------------------------------
\section{Acknowledgments}

\noindent{\bf Acknowledgments}:
This work is supported in part by the National Natural Science Foundation of China under Grants No. 11705034. and No. 12065006, and by the Project of GuiZhou Provincial Department of Science and Technology under Grant No. QKHJC[2020]1Y035.\\

\end{document}